\documentclass[preprint,pra,aps,nofootinbib,superscriptaddress]{revtex4-2}
\usepackage[mathscr]{euscript}
\usepackage{graphicx}
\usepackage{float,epsfig}
\usepackage{dcolumn}
\usepackage{bm}
\usepackage{graphicx}
\usepackage{amsmath,amssymb,amsthm}
\usepackage{subfigure}
\usepackage{color}
\usepackage[colorlinks=true,linkcolor=blue]{hyperref}
\usepackage[normalem]{ulem}
\usepackage[makeroom]{cancel}
\usepackage{caption}
\usepackage{subcaption}
\usepackage{textcomp}
\usepackage{ragged2e}
\usepackage{mathalfa}

\newcommand{\bea}{\begin{eqnarray}}
\newcommand{\eea}{\end{eqnarray}}
\newcommand{\beq}{\begin{equation}}
\newcommand{\eeq}{\end{equation}}

\def\/{\over}
\definecolor{purple}{rgb}{0.8,0,1}

\newcommand{\br}{{\bm r}}

\begin{document}

\title{Topological pumping of light governed by Fibonacci numbers}

\author{Ruihan Peng}
\affiliation{School of Physics and Astronomy, Shanghai Jiao Tong University, Shanghai 200240, China}

\author{Kai Yang}
\affiliation{School of Physics and Astronomy, Shanghai Jiao Tong University, Shanghai 200240, China}

\author{Qidong Fu}
\affiliation{School of Physics and Astronomy, Shanghai Jiao Tong University, Shanghai 200240, China}

\author{Yanli Chen}
\affiliation{School of Physics and Astronomy, Shanghai Jiao Tong University, Shanghai 200240, China}

\author{Peng Wang}
\affiliation{School of Physics and Astronomy, Shanghai Jiao Tong University, Shanghai 200240, China}

\author{Yaroslav V. Kartashov}
 \affiliation{Institute of Spectroscopy, Russian Academy of Sciences, Troitsk, Moscow, 108840, Russia}

\author{Vladimir V. Konotop}
 \affiliation{Departamento de F\'isica and Centro de F\'isica Te\'orica e Computacional, Faculdade de Ci\^encias, Universidade de Lisboa, Campo Grande, Edif\'icio C8, Lisboa 1749-016, Portugal}

\author{Fangwei Ye} 
\email{fangweiye@sjtu.edu.cn}
\affiliation{School of Physics and Astronomy, Shanghai Jiao Tong University, Shanghai 200240, China}
\affiliation{School of Physics, Chengdu University of Technology, Chengdu, China}
\date{\today}


\begin{abstract}
\textbf{Topological pumping refers to transfer of a physical quantity governed by the system topology, resulting in quantized amounts of the transferred quantities. It is a ubiquitous wave phenomenon typically considered subject to  exactly periodic adiabatic variation of the system parameters. Recently, proposals for generalizing quasi-periodic topological pumping and identifying possible physical settings for its implementation have emerged. In a strict sense, pumping with incommensurate frequencies can only manifest over infinite evolution distances, raising a fundamental question about its observability in real-world finite-dimensional systems. Here we demonstrate that bi-chromatic topological pumping with two frequencies, whose ratio is an irrational number, can be viewed as the convergence limit of pumping with two commensurate frequencies representing the best rational approximations of that irrational number. In our experiment, this phenomenon is observed as the displacement of a light beam center in photorefractive crystals induced by two optical lattices. The longitudinal periods of the lattices, that in the paraxial approximation emulate two pumping frequencies, are related as Fibonacci numbers, successively approaching the golden ratio. We observed that a one-cycle displacement of the beam center at each successive approximation is determined by the relation between successive Fibonacci numbers, while the average direction of propagation (emulating average pumping velocity) of the beam is determined by the golden ratio.
}
 
\end{abstract}
\maketitle

\section*{1 Introduction} 
The fundamental significance of the  topological (alias Thouless) pumping lies in its dependence on the global, rather than local, properties of a system. Typically such pumping is investigated over one (or several) cycles of the evolution of the system parameters and is characterized by quantized amounts of the transported quantities, hence it is also known as a quantized transport.
Ubiquity in wave systems and topological protection, making it insensitive to the local imperfections of the medium, are among the reasons for growing interest in this phenomenon. Predicted by Thouless~\cite{Thouless1983} for electrons in a periodic potential, quantized transport has been experimentally implemented in diverse physical systems~\cite{Citro23}, featuring periodicity of the propagation medium and cyclic evolution of the parameters. The phenomenon was experimentally observed in photonic~\cite{Zilberberg2018, Wang2022, Benalcazar2022, Ke2016, Cheng2022}, atomic \cite{Lohse2016, Nakajima2016, Taddia2017, Lohse2018}, spin~\cite{Ma2018}, mechanical \cite{Rosa2019}, and acoustic~\cite{Cheng2020,Chen2021} systems. 
Topological pumping can also be implemented and observed in aperiodic systems, like quasi-periodic arrays of optical waveguides~\cite{Kraus2012, Verbin2015, Kraus2013} and photonic crystals~\cite{Yang2024}, disordered \cite{Cerjan2020} lattices, and  fermions in a quasi-periodic potential~\cite{Nakajima2021}.  


Quantized transport has its own unique features and characteristics, even though it is conceptually similar to other topological phenomena, such as the celebrated quantum Hall effect~\cite{Thouless1982}. In particular, quantized transport manifests itself in the time-coordinate (1+1)D space (in the paraxial optics, the role of time is played by the propagation distance). Observation of such quantized transport usually requires adiabatic and periodic in time variation of the parameters of a medium. Adiabatically, in the first place allows avoiding undesirable inter-band transitions~\cite{Avron1987}, while the cycling defines the temporal interval over which pumping is quantized. Respectively, experimental detection of the dependence of the transported quantity on the global properties of the system requires tracking the transport over at least one cycle (in some cases, half a cycle~\cite{Fu2022a}) of as slow variation of the medium, as possible.

In all studies of quantized transport mentioned above, the dynamical potentials or lattices leading to transport of excitations were aperiodic in space (in atomic systems) or in the transverse direction (in optical systems), while remaining periodic in time or along the evolution direction, respectively. However, principally new phenomena may arise when the system is modulated at two different frequencies \cite{Schiavoni2003, Gommers2005, Grossert2016, Gorg2019, Minguzzi2022}, especially when they are incommensurate. We mention interplay between quasi-periodic driving and spatial randomness in various physical systems~\cite{Hatami2016, Zhao2022}, the formation of long-lived phases of matter~\cite{Long2021, Else2020, Nathan2021}, and Floquet engineered topological states~\cite{Qi2021, Kolodrubetz2018, Nathen2021, Hu2024}. All these studies were not concerned with the quantized transport and dealt with discrete systems, i.e., different from the setting with light pumping considered here. Practical feasibility of observing a quantized transport over a finite time (or length) in real samples is severally restricted by at least two factors: necessity of using extremely (theoretically infinitely) long propagation distances (long samples) and by lacking of a well-defined period of modulation over which the quantization of transport can be defined. Thus, experimental observation of the fundamental phenomenon of topological pumping enabled by quasi-periodic driving remains elusive.

In this work, we develop theoretical description and report the first experimental observation of the topological pumping under quasi-periodic bi-chromatic pumping, i.e. under principally new conditions, when dynamical potential causing transport is quasi-periodic in evolution variable. We thus explore the propagation of a paraxial light beam in a photorefractive crystal, which is exactly periodic in the transverse direction and quasi-periodically modulated along the direction of the light propagation. The system governed by the linear Schr\"odinger equation. To observe pumping that in our case occurs for bulk states we invoke the theory of the best rational approximations (BRA)~\cite{Khinchin}, which has been successfully applied in previous studies of systems with a quasi-periodic potentials and periodic pumping (see e.g.~\cite{Ostlund1983, Diener01, Modugno09, Marra2020, Zezyulin2022, Yang2024}). In experiment we induce in photorefractive crystals a series of longitudinally periodic lattices approaching longitudinally quasi-periodic lattice with progressively increasing accuracy. This allows us to work in quasi-adiabatic regime to overcome the constraint imposed by the absence of a well-defined cycle of evolution, and thus use at each approximation the results of the conventional Thouless theory.

We have observed that when longitudinal frequencies of sublattices forming the structure are related as Fibonacci numbers (i.e., in the limit approximate the golden ratio), a one-cycle displacement of the beam center at each successive approximation is also determined by the relation between successive Fibonacci numbers, what has been confirmed by theory developed for the paraxial light bean propagating in our structure. With increase of the number of approximation, upon convergence to authentic quasi-periodic system, the average direction of beam propagation is found to approach the golden ratio conjugate.

\section*{2 Results}  

\subsection*{2.1 Periodic approximants for the quasi-periodic pumping}
 
The propagation of a slit (i.e. uniform in the $x$ direction) low-power extraordinary polarized probe light beams in a quasi-periodic optically induced lattice is described by the dimensionless paraxial ({\em alias} Schr\"odinger) equation
\begin{align}
\label{nlse}
    i\partial_ z \Psi=H_\varphi (y, z)\Psi
\end{align}
for the complex amplitude of the electric field $\Psi$ of the probe beam. The Hamiltonian is given by $H_\varphi (y, z)=-(1/2)\partial_y^2+V_\varphi (y, z)$, where $y$ and $ z$ are the dimensionless transverse and longitudinal (evolution) coordinates.  The "dynamical" optical lattice potential
\begin{align}
\label{potential}
    V_\varphi(y,z)\equiv V(y,\eta,\zeta)= \frac{V_0}{1+I(y,\eta,\zeta)}
\end{align}
with
\begin{align}
\label{Iy}
 I(y,\eta,\zeta)=4p_1^2|e^{-iy}\cos y +e^{-i\eta/2}\cos(\eta/2) |^2+
4p_2^2\cos^2(\zeta/2)
\end{align}
$\eta=y-\alpha z$, and $\zeta=y-\alpha z/\varphi$, is created by superposition of three sublattices created by ordinary polarized beams: a reference sublattice and two sublattices sliding with angles $\alpha$ and $\alpha \varphi^{-1}$ with respect to the reference one. This form of the potential is determined by the mechanism of the response of the photorefractive crystal to external illumination, when ordinary polarized lattice-creating beams propagate in linear regime, but nevertheless create inhomogeneous refractive index landscape for extraordinary polarized probe beam $\Psi$. The sublattice depths $p_1^2$ and $p_2^2$ are proportional to the intensities, $E_{1,2}^2$, of the lattice-creating beams (see \textbf{Methods}). In the context of the Thouless pumping the sliding angles emulate the frequencies of the driving (or alternatively, the velocities of motion of the lattices) while $ z$ plays the role of time in the Schr\"odinger equation. In our model the dimensionless units are chosen such that $V_\varphi (y, z)$ is $2\pi$-periodic with respect to $y$, denoted by $Y=2\pi$ in the following, while two driving (dimensionless spatial) frequencies are equal to $\alpha$ and $\alpha \varphi^{-1}$. Notice that the adiabaticity of the pumping process requires $0<\alpha\ll 1$ (see \textbf{Methods}). We consider linear evolution of $\Psi$ subject to the condition that the probe beam has sufficiently low intensity.

\begin{figure}[H]
     \centering
     \includegraphics[width=1\linewidth]{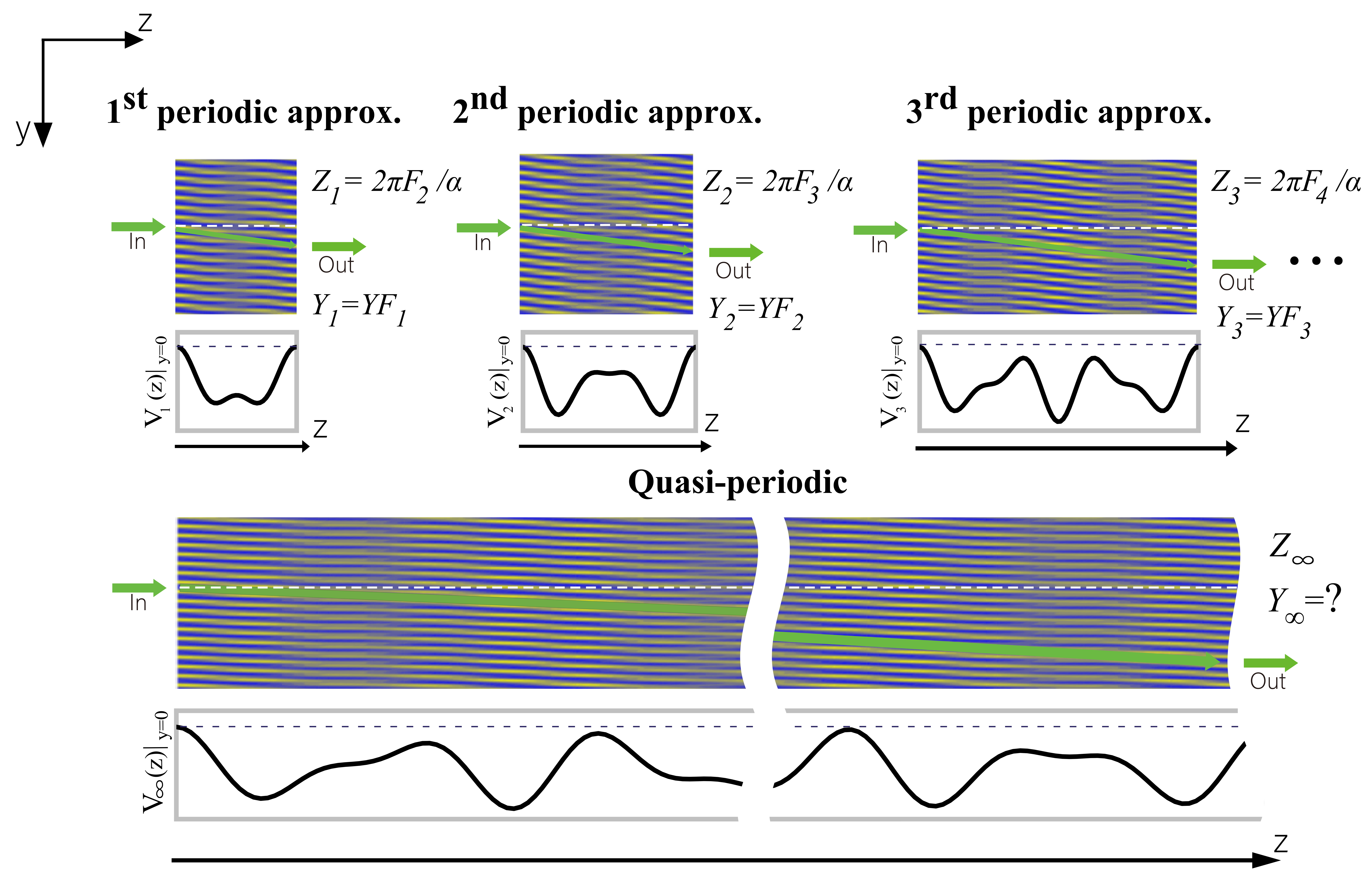}
     \captionsetup{font={stretch=1}}
     \captionsetup{labelfont=bf}
     \caption{
     {\bf 
     Conceptual schematic of periodic approximants illustrating light pumping in quasi-periodic lattices.} 
    The first four periodic approximants, characterized by the optical potential $V_{n}(y,z)$, and the quantized displacement $Y_n$ of the output beam for the localized input beam after one pumping period $Z_n$. The $z$-periods $Z_n$ are depicted in proportion to each other, i.e. with $Z_1:Z_2: Z_3=1: 1.5: 2.5$. The green line indicates the direction of the light pumping identified by the trajectory of beam center of mass. The goal of the study is to predict the light displacement  $y_c$ in the limit $n\to\infty$ (bottom panel), $Y_n$, and to experimentally approach this limit using its longitudinally periodic approximants (upper three panels). The black curve $V_n(z)|_{y=0}$ depicts the lattice profile varying over $z$, following the white dashed line at $y=0$ of $V_n(y, z)$. The purpose of the study is to obtain the pumping velocity, defined by $Y_\infty/Z_\infty$, in the authentic quasi-periodic lattices.
    }
     \label{fig1}
 \end{figure}

Optical potential $V_\varphi$ is quasi-periodic in $z$ when the parameter $\varphi$ is the irrational number, i.e. when two driving frequencies $\alpha$ and $\alpha \varphi^{-1}$ are incommensurate. Our consideration is detailed to target the golden ratio conjugate $\varphi^{-1}=(\sqrt{5}-1)/2$, but this does not restrict the generality of the approach, which allows direct generalization to any other irrational relation between the frequencies. Namely, we use the fact that any irrational number can be represented by a continued fraction whose convergents are its best rational approximations (BRAs) of the second kind~\cite{Khinchin}. In the standard notations, the golden ratio conjugate (defining the driving frequency $\alpha \varphi^{-1}$)  is written as $(\sqrt{5}-1)/2=[0;1,1,1,..]$ and has the BRAs  given by the quotients $ \left\{F_0/F_1,F_1/F_2,F_2/F_3, \cdots\right\}$ of the Fibonacci numbers $F_n$. Thus, $F_n/F_{n+1}$ is  $n$th BRA for the golden ratio conjugate and  $\varphi^{-1} =\lim_{n\to\infty} (F_{n}/F_{n+1})$. Since any two subsequent Fibonacci numbers $F_n$ and $F_{n+1}$ are coprime integers, the optical potential $V_n(y,z)=V (y, \eta, \zeta_n)$ with $\zeta_n=y- (F_n/F_{n+1})\alpha z$, corresponding to the approximation of $\varphi^{-1}$ by the ratio $F_n/F_{n+1}$, results in pumping using two-frequencies that leads to (longitudinal in our case) period of $Z_n=2\pi F_{n+1}/\alpha$. This observation suggests that instead of studying directly the light propagation in the system governed by the Hamiltonian $H_\varphi(z)$ one can explore the propagation governed by its periodic approximants $H_n (z)=-(1/2)\partial_y^2+V_n(y,z)$. Thus, the approximants are obtained by the replacement of the golden ratio conjugate $\varphi^{-1}$ with its $n$th BRAs $F_n/F_{n+1}$. The schematics of this approach is presented in Fig.~\ref{fig1} and it shows how increasing the number of BRA (as long as this is consistent with available sample length due to increasing $Z_n$ and requirements of adiabaticity) allows observation of gradual transition to quasi-periodic pumping.

\subsection*{2.2 Beam center displacement upon pumping}

When applying the method of periodic approximants (instead of considering a longitudinally infinite system) to describe quasi-periodic bi-chromatic pumping, one must address the following question: Does this limit exist, and if so, can it be achieved with sufficient accuracy in a finite physical system? To demonstrate that the answer is positive, we consider the displacement of the center of mass (COM) $y_n(z)=\int y |\Psi_n|^2dy/\int{|\Psi_n|^2dy}$ of the light beam after one period of the adiabatic evolution governed by $n$-th approximant, i.e., by the Schr\"{o}dinger equation (the respective field is denoted by $\Psi_n$)

\begin{align}
\label{SEn}
 i\partial_z \Psi_n=H_n (z)\Psi_n.   
\end{align}

Following the standard approach, one calculates the instantaneous spectra of the approximants, i.e., the propagation constants $\beta_{\nu k}^{n}(z)$, by solving $H_n \phi_{\nu k}^{n}=-\beta_{\nu k}^{n}\phi_{\nu k}^{n}$, where $\phi_n^{\nu k} (y,z)=e^{iky} u_{n}^\nu(y, z,k)$ with $u_{n}^\nu(y, z,k)=u_{n}^\nu(y+2\pi, z,k)$, are the respective Bloch functions, $\nu$ and $k$ denote the Bloch band and wavenumber in the reduced Brillouin zone. According to the Thouless theory~\cite{Thouless1983}, a relatively narrow probe beam, whose spectral width coincides with the width of the $\nu$-th band of the Hamiltonian $H_n$, after one pumping cycle $Z_n$  experiences the quantized displacement $Y_n^\nu=Y C_n^\nu$ where $Y_n^\nu=y_n(Z_n)$, $C_n^\nu=(i/2\pi)\int_{0}^{ Z_n}d z\int_{-1/2}^{1/2} dk  \Omega_n^\nu (k, z)$  is the Chern index of the excited band $\nu$, and $\Omega_n^{\nu} (k, z)=\left\langle \partial_ z u_{n}^\nu|\partial_k u_{n}^\nu\right \rangle-\left\langle \partial_k u_{n}^\nu |\partial_ z u_{n}^\nu\right\rangle
$ is the Berry curvature ($\langle f|g\rangle=\int_0^{2\pi} f^*g dy$). 

Considering approximants of sufficiently large  $n$, such that $F_{n+2}\gg 1$, in the interval $ z\in [0,   Z_n]$ one can expand the difference $V_{n+1}-V_{n}$ in the Taylor series, to represent the $(n+1)$th approximant in the form $H_{n+1}=H_{n}+W_n(y, z)$, where  
\begin{align*}
	W_n(y, z)=   \frac{(-1)^n \alpha z }{F_{n+1}F_{n+2}} \frac{\partial V_n}{\partial\zeta_n}+\mathcal{O}\left( \frac{V_n}{F_{n+2}^2} \right)=o\left( V_n\right)
\end{align*}
Here we used the property of the Fibonacci sequence $F_{n+2}F_n-F_{n+1}^2=(-1)^{n+1}$, the estimate $\alpha  z<\alpha Z_n
\ll F_{n+1}F_{n+2}$, as well as the assumption about smoothness of the potential, expressed by $|\partial V_n/\partial\zeta_n|\sim |V_n|$. Thus, under the above conditions the Hamiltonian $H_{n+1}$  can be viewed as a perturbation of the the $n$-th approximant. This allows us to employ the perturbation theory to obtain the relation $C^\nu_{n+1}=C^\nu_{n} +C^\nu_{n-1}	+\mathcal{O} (1/F_{n+2})$ (see \textbf{Methods}). Since $C_n^\nu$ are integers, while the small asymptotic term $\mathcal{O} \left(1/F_{n+2}\right)$ is not, the exact relation $C^\nu_{n+1}=C^\nu_{n} +C^\nu_{n-1}$ must hold.  This indicates that the Chern indices, derived using the employed method for constructing approximations of quasi-periodic driving, follow the Fibonacci sequence rule. While this sequence was obtained for large positive integers $n$, we conjecture (and check numerically, see below) its validity for all $n\geq 0$. Then, considering that $V_{n=0}=V(y,y-\alpha z, y)$ and $V_{n=1}=V(y,y-\alpha z, y-\alpha z)$, we obtain for the upper excited band $\nu=1$ the respective approximants $C^{1}_{0}=0=F_{0}$ and $C^1_{1}=1=F_{1}$, what implies that $C^1_n=F_n$ for all positive integers $n$. This leads to our main result: the velocity of the quantized pumping governed by periodic approximants, i.e., 
\begin{align}
\label{v_n}
    v_n=\frac{Y C^1_n}{Z_n}
\end{align}
(recall that the transverse period is $2\pi$), in the limit $n\to \infty$ approaches the velocity of the respective bi-chromatic quasi-periodic pumping:
\begin{align}
    \label{v_aver}
    v=\lim_{n\to\infty}v_n=
    \alpha\lim_{n\to\infty}\frac{F_n}{F_{n+1}}=\frac{\alpha}{\varphi}
\end{align}
which is determined by the ratio between the frequencies, i.e. by the golden ratio $\varphi$ in our case.

\subsection*{2.3 Dynamics of topological pumping}

\begin{figure}[H]
     \centering
     \includegraphics[width=\linewidth]{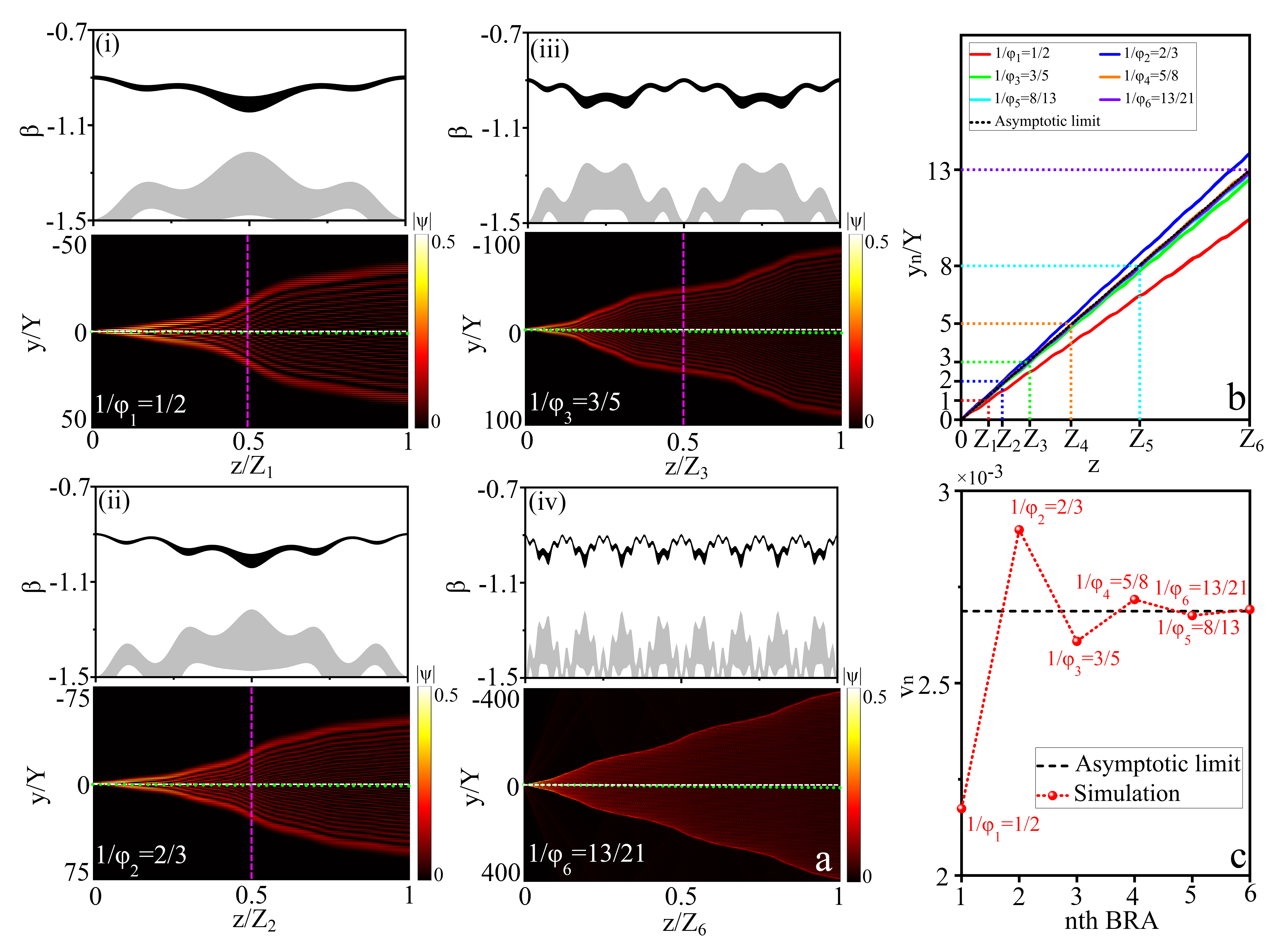}
     \captionsetup{font={stretch=1}}
     \captionsetup{labelfont=bf}
     \caption{
     {\bf Quantized transport of light in the lowest periodic approximants for $\varphi^{-1}=(\sqrt{5}-1)/2$.} 
     ({\bf a}) The first (black) and the second (gray) bands of the spectra of the first three approximants (i, ii, iii), studied experimentally (see Fig.~\ref{fig3} below) and for the 6-th  BRA (iv) at which the one-cycle displacement closely approaches the quasi-periodic limit. The vertical magenta dashed lines in (i, ii, iii) indicate the propagation distances $Z_n/2$ at which the output beam was observed in the experiment. The green dots  mark the positions of the COM, while white dashed line indicates the $y=0$ position, where the incident beam was launched. The  corresponding diffraction of the light beam over single periods $Z_n$ is shown in the lower panels.  
     ({\bf b}) The dependence of the displacement $y_n$ of the beam COM normalized to the transverse lattice period $Y=2\pi$ versus propagation distance $z$ for the first six BRAs. The asymptotic limit prediction (Eq.~(\ref{v_aver}) from the main text) is shown by the dashed line.
     ({\bf c}) The average velocity $v_n$ (the band index $\nu=1$ is omitted here) of the COM displacement for different approximants are shown as the red dots converging with increase of $n$ to the asymptotic limit $ \alpha/\varphi$, shown with dashed black line. In all simulations, $V_0=-2.5$, that corresponds to the voltage 600~V applied to the sample in experiment.
     }
     \label{fig2}
 \end{figure}

 The instantaneous borders of the allowed bands versus propagation distance $z$ for some lowest approximants, described by Eqs.~(\ref{potential}) and (\ref{Iy}) are presented in Fig.~\ref{fig2}(a) (the borders of the bands for even higher orders of approximation are presented in Fig. S12 in \textbf{Supplementary Information}). While for any approximant the gaps remain open for all distances $z$, the   particularly large gap between the upper and second bands, guarantees that the incident beam, initially exciting only the band $\nu=1$ (as it occurs in our experiments), will populate only this band during the entire evolution of the system. The Chern number $C^{\nu}_n$ for the lowest BRAs $n=1,..,6$  are computed to be  $1, 2, 3, 5, 8, 13$, respectively. As it was predicted above, this is the sequence of the Fibonacci numbers $F_1,...,F_6$, thus, confirming that the transport of light in successive periodic approximants is indeed governed by Fibonacci numbers. 

The outcome of direct simulations of pumping dynamics governed by Eq.~(\ref{SEn}) with an input Gaussian beam $\Psi_{\rm in} (y)=e^{-y^2/r_0^2}$ of width $r_0=2.2$, is presented in the lower panels of Fig.~\ref{fig2}(a). This input corresponds to a localized excitation that populates the entire $\nu=1$ band (consistent with the experimental excitation conditions). Indeed, such Gaussian beam represents an accurate approximation (see e.g.~\cite{Alfimov}) of the Wannier function of the upper band, which corresponds to a superposition of all Bloch states of this band with different Bloch momenta~\cite{Kohn}. Moreover, our results show that Thouless pumping can be observed even for broader beams with width up to $r_0=19.8$, see Fig. S7 and S8 and discussion in \textbf{Supplementary Information}. In all simulations, lattice amplitudes are $p_1^2=0.09$ and $p_2^2=0.49$, and we consider a small angle $\alpha \approx 0.004$, which defines the sliding rate of the two sublattices and corresponds to the best possible approach to  the adiabatic regime of lattice variations achievable in the experiment. All simulations use the same initial condition $\Psi_{\rm in}$ and  are carried out over distances corresponding to one pumping cycle. In Fig. \ref{fig2}(a) we show propagation dynamics for the first, second, third, and sixth orders of approximation, while dynamics for the fourth and fifth order approximations are presented in Fig. S12 of \textbf{Supplementary Information}. One can see from Fig. \ref{fig2}(a) that as the order of rational approximation increases, the $z$-period $Z_n$ of the optical potential increases as well, that leads to stronger diffraction broadening of the beam on one pumping cycle. Nevertheless, the displacement of the beam COM $y_n(z)$ after one pumping cycle increases with the order of approximation too. The dependence of the computed displacement of the beam COM $y_n(z)$ is presented in Fig.~\ref{fig2}(b) for approximations of different order $n$. Despite considerable diffraction of the beam, clearly visible in Fig.~\ref{fig2}(a), the displacement of the center of mass $y_n(z)$ is dictated by the topology of the system, in a sense that displacement over one $z$-period, $Z_n$, is quantized and is given by $Y_n^\nu=Y C_n^{\nu}$ for different BRAs [see dashed lines in Fig.~\ref{fig2}(b)]. Notably, the displacement after one cycle obtained numerically is verified to follow the recursion relation $Y_{n+1}^\nu=Y_n^\nu+Y_{n-1}^\nu$ [Fig.~\ref{fig2}(b)]. Thus, despite increasing with $n$ diffraction broadening of the beam, the one-cycle velocity of the pumping defined by (\ref{v_n}) and illustrated in Fig.~\ref{fig2}(c) by the red dots connected by the dashed red lines, deviates from theoretically predicted value (\ref{v_aver}) [black dashed line in Fig.~\ref{fig2}(c)] only for small orders of approximation $n$, and rapidly converges to (\ref{v_aver}) with increase of the order of BRA. Thus, studying pumping in different approximations of exact periodic bi-chromatic pumping allows for the observation of a relatively rapid transition to the average one-cycle pumping velocity. Furthermore, it confirms that such quasi-periodic pumping is inherently of topological origin.

\subsection*{2.4 Experimental observation of the convergence of the average displacement}

The above predictions based on the analysis of spectrum of the periodic approximants $V_n (y,z)$  and on modeling propagation of light beams in such optical potentials have been corroborated through two distinct experiments. The results of which are depicted in Fig.~\ref{fig3} and Fig.~S5 (see \textbf{Supplementary Information}). We employed the optical induction technique ~\cite{efremidis2002,fleischer2003} to generate (using interference of sets of ordinary polarized plane waves, see \textbf{Methods}) a sequence of periodic lattices within SBN:61 photorefractive crystal with dimensions $5\times5\times20 ~\textrm{mm}^3$. These lattices correspond to the BRAs with increasing  $n$ of quasi-periodic structures defined by two irrational numbers $\varphi^{-1}=\pm (\sqrt{5}-1)/2$ [introduced in Eq. (\ref{Iy})]. The maximal intensity of the lattice-writing beam was around 
$0.54~\textrm{mW}/\textrm{cm}^2$, while the intensity of the probe beam was approximately ten times lower than the lattice-writing beam to ensure its linear propagation regime. To examine light propagation in the induced optical lattices, we utilized a low-power, extraordinarily polarized probe laser beam. The beam was initially directed through a cylindrical lens to produce a quasi-one-dimensional slit profile, that is nearly uniform along the $x$-axis and that has a finite width of $30~\mu$m along the $y$-axis. This beam was then directed normally onto the front facet of the crystal, aligned with one of the local lattice maxima, and covered approximately one lattice stripe, as shown clearly in Fig.~\ref{fig3}(a). This is necessary for nearly uniform excitation of the $\nu=1$ band, while other higher bands remain unexcited. The beam’s propagation dynamics within the crystal was recorded using a $z$-scanning CCD camera (see \textbf{Methods} for details). 
Since in the experiment only the band $\nu=1$ is populated, to shorten notations we drop the band index $\nu$ in what follows.

Figures~\ref{fig3}(a)–(c) display the intensity distributions of the dynamical lattices corresponding to the 1st, 2nd, and 3rd BRAs of quasi-periodic structure within half of the longitudinal period $Z_n$. The lattice profile is uniform along the $x$ direction and periodic in both the $y$ and $z$ directions. The induced lattices are identical at $z=0$ and $z=Z_n$; however they vary differently with $z$, because of different driving frequencies $\alpha \varphi_n^{-1}$ at different BRAs. The higher-order lattices also exhibit more oscillations along $z$ within one $z-$period (or within half of $z$-period in the experiment), indicating their progressive convergence toward the final, truly quasi-periodic lattices. While in case of approximants the lattice is expected to reproduce itself  after a longitudinal period, in experiment, the propagation distance is selected to ensure that finite sample length of $20~\textrm{mm}$ is sufficient to accommodate half of the longitudinal period, namely, $Z_1/2\approx 8 ~\text{mm}$, $Z_2/2 \approx 12~\text{mm}$, and $Z_3/2 \approx 19 ~\text{mm}$ in physical units, respectively, at least in the 3rd BRA of the bichromatic quasiperiodic lattice. Note that, due to rapid increase of $Z_n$ with $n$, half of the period for 4th BRA already exceeds the sample length. Importantly, due to parity-time symmetry of $H_n(z)$, one finds that the Chern number calculated over half of the period amounts to half of the Chern number $C_n$ obtained over the whole period~\cite{Fu2022a}. This allows us to experimentally investigate the optical transport and COM displacement for the first three BRAs within a $2$ cm-long sample.

Figures~\ref{fig3}(d)–(f) illustrate the dynamics of the pumping process by showing experimentally measured intensity distributions of the beam in the $(y,z)$ plane in dynamical lattices depicted in Fig.~\ref{fig3}(a)–(c), while Fig.~\ref{fig3}(g) shows measured COM displacement $y_n(z)$ of the probe beam in the $n$th BRA. Green dots in Fig.~\ref{fig3}(d)–(f) illustrate the position of COM after each $1~\textrm{mm}$ of propagation superimposed on the intensity distribution in the $(y,z)$ plane and clearly show its transverse displacement in positive direction of the $y-$axis upon propagation. The measured dynamics closely resembles theoretical results shown in Fig.~\ref{fig2}. 
The differences between measured experimental intensity distributions and theoretical patterns can be attributed to the impact of weak drift and diffusion nonlinearity on weak probe beam and its possible back-action on the lattice that are always present in real photorefractive crystal, but are not accounted for in theoretical model.
Figure~\ref{fig3}(g) compares the experimentally measured COM displacement with theoretical prediction for different BRAs. Since the velocity of pumping is determined by the global topological properties of the bands (characterized by spatiotemporal Chern numbers), rather than by local imperfections or disorder inherently present in our experimental setup, the pumping process is expected to be robust against any local perturbations. This is what one observes in Fig. \ref{fig3}(g) that despite the presence of some dispersion in measured COM displacement shows good agreement with theoretical prediction from Eq. (\ref{v_aver}). Slight discrepancy between experimental COM measurements and theoretical predictions, which assume adiabatic conditions, can be attributed to non-adiabatic effects and potentially to weak nonlinear effects that are not accounted for in theory. Importantly, in our setting the voltage applied to the sample affects the depth of the optical potential, but it does not affect its topological characteristics (i.e. Chern numbers of the bands) and therefore it should not affect the rate of pumping determined by global topological properties of the lattice. Indeed, increasing voltage leads to gradual suppression of diffraction, as we demonstrated in Fig.~S2 in \textbf{Supplementary Information} by comparing intensity distributions at $400$, $600$ and $800~\textrm{V}$. At the same time, the velocity $v_n$, plotted in Fig.~\ref{fig3}(h) for different applied voltages remains practically independent of this parameter, confirming topological nature of this characteristic. Fig.~\ref{fig3}(h) illustrates also our main observation - gradual convergence of the rate of observed pumping (defined using COM position at the distance $Z_n/2$) to asymptotic value corresponding to pumping in truly quasi-periodic lattice with increase of the number of a BRA. These experimental results strongly support the conclusion about robustness of pumping to variations in lattice parameters and to disorder and about topological nature of this phenomenon.

Similar observations are also reported in the \textbf{Supplementary Information} for another example of the bi-chromatic pumping, characterized by the irrational number $\varphi^{-1}=-(\sqrt{5}-1)/2$. This corresponds to opposite velocities of two sliding lattices.  Notably, the Chern numbers of the approximants in this case change their signs compared to those shown in Fig.~\ref{fig3}, while preserving their absolute values. As a result, the center of mass (COM) of the light beam shifts in the negative $y$-direction. The results are presented in Fig.S1, S3, S4 and S6.

\section*{3 Discussions}

The approach based on periodic approximants, along with its experimental implementation reported here, provides a unique physical interpretation of quantized pumping under quasi-periodic bi-chromatic driving. This interpretation characterizes quasi-periodic pumping by the velocity, which is viewed as the limit of quantized velocities measured after one cycle of each periodic bi-chromatic driving, where the periods are related through the BRAs of the given irrational number. Thus, quasi-periodic pumping is intrinsically linked to the topological properties of the systems in which it is observed. The mentioned limit is closely approximated even with approximants of relatively low orders. From broad perspective, this study shows that topological robustness of the pumping process persists even in systems driven at incommensurate frequencies, challenging the traditional requirements for strict periodicity of driving.

The reported transport therefore represents a general topological phenomenon, as the derived results do not rely on the specific type of physical system, the local characteristics of the transverse lattice, or the dependence of pumping on the evolution coordinate. Therefore, one can expect that the described quasi-periodic driving can be implemented and observed in physical systems of other nature, like systems of cold atoms~\cite{Lohse2016, Nakajima2016, Lohse2018, Nakajima2021} and plasmonics~\cite{Fedorova2020}. Our results open up a new pathway for designing and acquiring topological structures/materials.

There are several open general questions regarding the physics of quasi-periodic pumping that are yet to be explored. In our experiments, the observed light beams undergo significant diffraction during pumping process. Its suppression requires substantial increase in the lattice depth, that is nevertheless limited by the capabilities of optical induction technique. A potential extension would be utilization of discrete waveguide arrays, where large waveguide depth can lead to substantial suppression of diffraction broadening~\cite{Ke2016}, or utilization of nonlinear effects that could suppress diffraction, enabling the pumping of a localized wavepacket~\cite{Fu2022a, Fu2022}. Another natural extension is the study of quasi-periodic bi-chromatic pumping in quasi-periodic lattices that support the propagation of localized, non-diffractive beams~\cite{Yang2024}. Novel transport scenarios can also be anticipated in two-dimensional settings, both in periodic lattices~\cite{Wang2022} and in moir\'e lattices, where light diffraction can be significantly suppressed~\cite{Wang2020}.  
 
From an experimental perspective, the setting described here provides a versatile platform for exploring the interplay between topology, symmetry, and, in the nonlinear regime, many-body effects. This could lead to the discovery and experimental validation of exotic phases of matter with unique transport properties and unveil new opportunities for engineering innovative quantum devices with unique transport properties.

\section*{4 Methods}

\subsection*{4.1 Experimental setup} 

As depicted in Fig.~\ref{fig.4}, the experimental setup utilizes the optical induction technique to create the lattice. Specifically, two continuous-wave, frequency-doubled Nd:YAG lasers operating at the wavelength of 532 nm, ordinarily polarized after passing through a half-wave plate (HWP), were employed to inscribe superlattices in a biased photorefractive crystal (SBN:61) with the refractive index $n_e\approx 2.2817$. The crystal, with dimensions of $5\times 5\times20~\text{mm}^3$, was oriented such that its optical axis was aligned with the $20~\text{mm}$ direction, indicated by the dashed line in beam path (1). To investigate the light propagation dynamics within the induced superlattice, an extraordinarily polarized beam at a wavelength of 632.8 nm, generated by a He-Ne laser, was directed into the sample (as shown in beam path (2)). A translation stage equipped with an imaging lens (L3) and a mounted CCD camera enabled the step-by-step recording of the induced optical lattices and the intensity of the probe beam, at intervals of 1 mm throughout the sample.

\subsection*{4.2 Optical potential} 

In our experiment, we interfere plane waves and imprint the resulting interference pattern of light into the photorefractive crystals SBN. The lattice-writing beams, after being expanded by a spatial filter (SF), are directed through an amplitude mask, Mask1, which has three pinholes symmetrically positioned along a line at  $y_{1,2,3}=-a, 0, a$, as shown in the inset of the Fig. S1. Note that here we use physical variables (their relation to the dimensionless ones used in the main text are given at the end of this subsection). The two beams originating from pinholes 2 and 3 are attenuated (AT) so that their amplitude is reduced to half, and then interfere with the beam originating from pinhole 1 (which remains unattenuated with an electric amplitude $E_1$) after being refracted by the lens $L_1$. The resultant interference intensity $I_1(\br)$ is given by $I_1 =(E_1^2/4)| 2e^{i \bm{k}_1  \cdot \br } + e^{i \bm{k}_2 \cdot \br } + e^{i \bm{k}_3 \cdot \br} |^2$, where $\br=(y, z)$, and $\bm{k}_m$ ($m=1,2,3$) is the wavevector of the $m$-th plane wave originating from the $m$-th pinhole, each having a transverse (along $y$-axis) and longitudinal (along $z$-axis) components,   $\bm{k}_1=( k_y, (k_0^2 - k_y^2)^{1/2} )$, $\bm{k}_2= (0, k_0)$, $\bm{k}_3= ( -k_y, (k_0^2 - k_y^2)^{1/2})$. In this way, one obtains a superposition of a stationary lattice with a sliding lattice, i.e., an optical potential emulating Thouless’s original setting~\cite{Thouless1983}. Such lattice has a transverse period $Y=2\pi/k_y=29.9\mu$m, and a longitudinal period $Z_1=4n_ek_0\pi/k_y^2=7.7~$mm. In order to add a second sliding lattice with the same transverse, but a different longitudinal period $Z_2$ related to the first one by the irrational number $\varphi$: $Z_2/Z_1=\varphi$, in addition to the Mask 1, we add a Mask 2 that features two pinholes at the position $y_{4, 5}=(1/\varphi\pm 1)a/2$. This creates the second sliding-lattice-writing beam with the longitudinal period, $Z_2=4n_ek_0\varphi \pi/k_y^2$.
The second sliding-lattice-writing beam is added incoherently thus producing the intensity distribution:

\begin{align}
\label{met1}
I_\varphi (y,z)
=\frac{E_1^2}{4}|2e^{i\bm{k}_1\cdot\br}
+e^{i\bm{k}_2\cdot\br}+e^{i\bm{k}_3\cdot\br}|^2+\frac{E_2^2}{4}|e^{i\bm{k}_4\cdot\br}+e^{i\bm{k}_5\cdot\br}|^2.
\end{align}

Here $\bm{k}_4=(\varphi k_y/2,(k_0^2-\varphi^{2}k_y^2/4)^{1/2})$, $\bm{k}_5=((\varphi^{-1}-1)k_y/2,(k_0^2-(\varphi^{-1}-1 )^2k_y^2/4)^{1/2})$. Note the intensity of the second sliding-lattice-writing-beam $E_2^2$  can differ from $E_1^2$ such that the photonic lattice induced by these beams could feature different modulation depths.  


The potential amplitude $V_0$ is controlled by the voltage $V$ added to the crystal obeying the empirical formula $ V_0=-k_0^2  n_e^4 r_{33} V/2 k_y^2 d $ ~\cite{Lederer2008}. Here $r_{33}=450~ \text{pm} \text{V}^{-1}$ is the electro-optic coefficient, $d$ is the transverse dimension of the sample ($d=5~$mm for our sample). The dimensionless variables are now introduced by replacements: $k_yy\to y$,  $ k_yz\to z$.  In these variables the magnitude of the sliding angle is given by $\alpha=k_y/n_ek_0$.

\subsection*{4.3 Perturbation theory for approximants}

Let us introduce approximants $H_{n, k}\equiv- ({1}/{2})\left(\partial_y+ik\right)^2-V_n(y,z)$ resulting in the eigenvalue problems $H_{n,k}u_n^\nu= \mu_n^\nu (k,z) u_n^\nu$ defined for integers $n$, 
and compute
\begin{align}
\label{CC}
\frac{C^\nu_{n}}{Z_n}-\frac{C^\nu_{n-1}}{Z_{n-1}}
     =\frac{i}{2\pi }\left(\frac{1}{Z_{n}}-\frac{1}{Z_{n-1}}\right)
     \int_{0}^{Z_{n}}dz\int_{-1/2}^{1/2} dk
 \Omega^{\nu}_{n}(k,z) 
 \nonumber \\
 +\frac{i}{2\pi Z_{n-1} }\int_{Z_{n-1}}^{Z_{n}}dz\int_{-1/2}^{1/2} dk
 \Omega^{\nu}_{n}(k,z) 
 +\frac{i}{2\pi Z_{n-1}}\int_{0}^{Z_{n-1}}dz\int_{-1/2}^{1/2} dk
		\left[ \Omega^{\nu}_{n}(k,z)-\Omega^{\nu}_{n-1}(k,z)\right] 
\end{align}
 Here following~\cite{Marra2020} we explore $C^\nu_{n}/Z_n$ which for our purposes is a more convenient quantity that the Chern numbers themselves.
 
Denoting $\epsilon=1/F_{n+2}\ll 1$ (to shorten notations) we can look for $u^\nu_{n}(y,z,k)$ in the form of perturbation expansion
 $u^\nu_{n}=u^{\nu}_{n-1}+ \epsilon v_{1}+ \cdots,$ and  $\mu^\nu_{n}=\mu^\nu_{n-1}+\epsilon \mu_1+... $. Thus for $z\in[0,Z_n]$ we have  
$ \Omega^{\nu}_{n-1}=\Omega^{\nu}_{(n-1)\pm 1} +\mathcal{O}(\epsilon) $. This, as well as the relation
\begin{align}
    \label{ZZ}
    Z_{n}= Z_{n-1}+ Z_{n-2}
\end{align}
which follows from the property of the Fibonacci sequence $F_{n+1}=F_{n}+F_{n-1}$, allow to recast (\ref{CC}) in the form
\begin{eqnarray*}
\label{CC2}
\frac{C^\nu_{n}}{Z_n}-\frac{C^\nu_{n-1}}{Z_{n-1}}
     =
      \left(\frac{1}{Z_{n}}-\frac{1}{Z_{n-1}}\right)C^\nu_{n}+\frac{i}{2\pi Z_{n-1} }\int_{Z_{n}}^{Z_{n-2}+Z_{n-1}}dz\int_{-1/2}^{1/2} dk\Omega^{\nu}_{n-1}(k,z)+\mathcal{O}(\epsilon) 
\end{eqnarray*}
Further simplification is achieved by using (\ref{ZZ}),  as well as the fact that $\Omega^{\nu}_{n-1}(k,z)$ is $Z_{n-1}$-periodic:
\begin{eqnarray}
\label{CC1}
\frac{C^\nu_{n}}{Z_n}-\frac{C^\nu_{n-1}}{Z_{n-1}}
     =-
      \frac{Z_{n-2}}{Z_{n}Z_{n-1}}C^\nu_{n}+\frac{C^\nu_{n-2}}{Z_{n-1} }+\mathcal{O}(\epsilon) 
\end{eqnarray}
After rearrangement of the terms, and using (\ref{ZZ}) once again we obtain $	C^\nu_{n+1}=C^\nu_{n} +C^\nu_{n-1}	+\mathcal{O} (1/F_{n+2})$ which is used in the main text.

\section*{Acknowledgments} 

V.V.K was supported by the Portuguese Foundation for Science and Technology (FCT) under Contracts UIDB/00618/2020 (DOI:10.54499/UIDB/00618/2020) and PTDC/FIS-OUT/3882/2020 (DOI:10.54499/PTDC/FIS-OUT/3882/2020). Y.V.K. acknowledges funding from the project FFUU-2024-0003 of the Institute of Spectroscopy of the Russian Academy of Sciences. R. P., K. Y., Q. F., Y. C., P. W. and F. Y. acknowledge funding from the Shanghai Outstanding Academic Leaders Plan (No.20XD1402000) and Shanghai Leading Talent Program 1 of Eastern Talent Plan (The 16th Shanghai Leading Talent Program). Q.F. and P. W. acknowledge the support of the National Natural Science Foundation of China (No.12404385, No.12304366) and China Postdoctoral Science Foundation (No.BX20230217, No.BX20230218, No.2023M742295, No.2024M751950).

\section*{Competing interests} 
The authors declare no competing interests.

\section*{Availability of data and materials}
The data that support the findings of this study are available from the corresponding author upon reasonable request.

\section*{Author contributions} 
R.P., K.Y. and Q.F. contribute equally to this work. All authors contribute significantly to the work.


\newpage

 \begin{figure}[H]
     \centering
     \includegraphics[width=\linewidth]{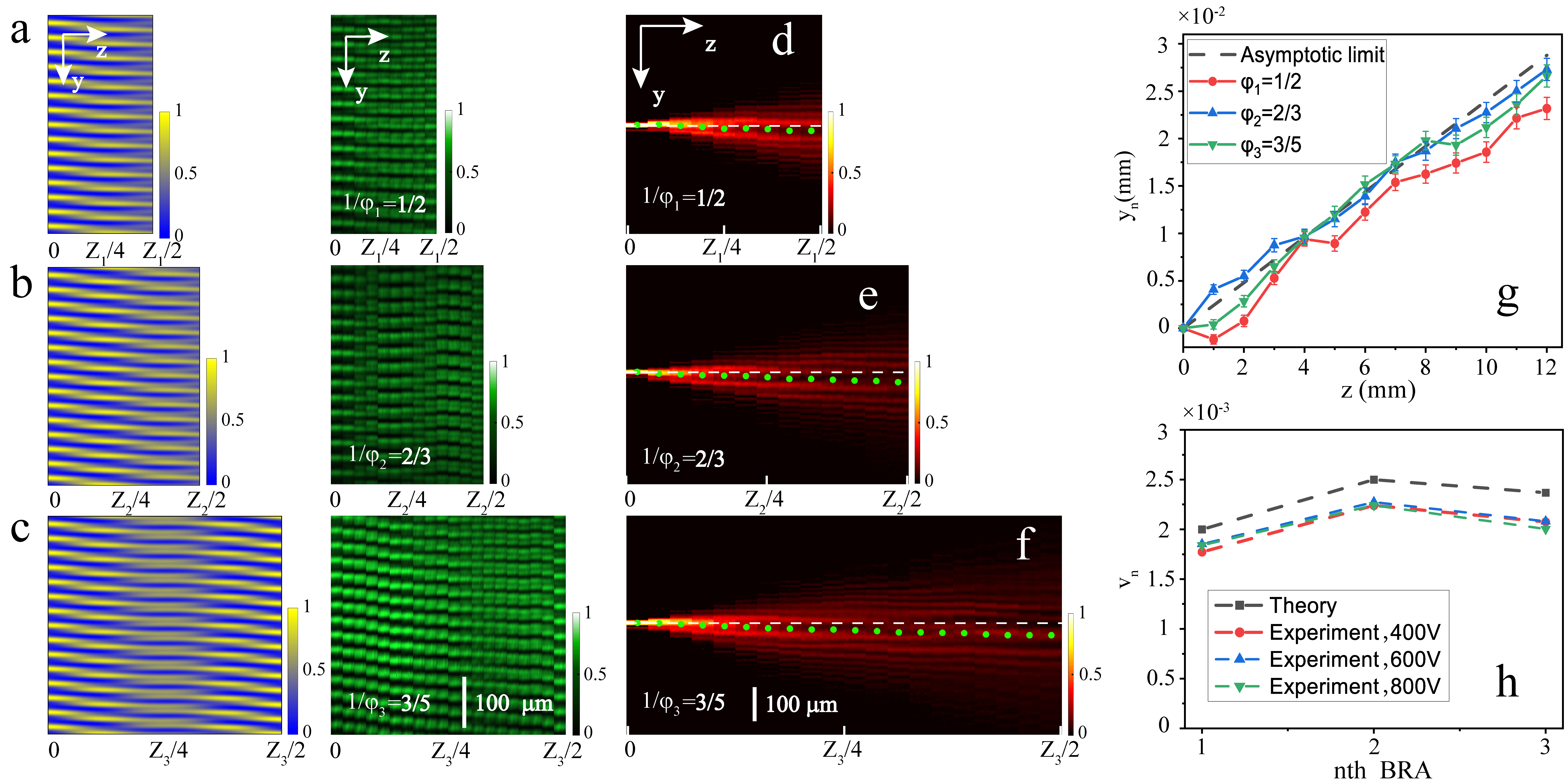}
     \captionsetup{font={stretch=1}}
     \captionsetup{labelfont=bf}
     \caption{
      {\bf Observation of light pumping in the approximants of a quasi-periodic bichromatic driving with $\varphi^{-1}=(\sqrt{5}-1)/2$.}
     (\textbf{a-c}) The profile of lattice corresponding to the three lowest BRAs $\varphi_1^{-1}=1/2$, $\varphi_2^{-1}=2/3$, and $\varphi_3^{-1}=3/5$ within half of the longitudinal period $Z_n$. The left column presents the simulation results, while the right column shows the experimentally induced optical lattices. The simulations are based on Eq.~\ref{met1}. Here, $k_0=1.18\times10^7$~m$^{-1}$, $k_y=2.1\times10^5$~m$^{-1}$, $E_1^2=0.19$~mW/~cm$^{2}$ and $E_2^2=0.35$~mW/~cm$^{2}$. (\textbf{d-f}) The intensity distribution of the probe beam in the $(y, z)$ plane, measured with the step of $1$ mm along the $z$-axis at the voltage of 600 V applied to the photorefractive crystal. The green dots superimposed on the plots showing the dynamics, mark the positions of the COM after each $1$ mm of propagation, while white dashed line indicates the $y=0$ position, where the incident beam was launched. Note that propagation distances in each case are shown in proportion to each other, with half of the pumping cycle being $Z_n/2\approx 8, 12, 19~\text{mm}$, respectively. (\textbf{g}) Experimentally measured COM position {\it versus} distance $z$ (dots) compared with the analytically predicted asymptotic behavior given by Eq.~(\ref{v_aver}) and shown by the dashed line. (\textbf{h}) The experimentally measured pumping rate, $v_n$, under different voltages (colored lines) compared with the theoretical prediction 
     (\ref{v_n}), with $\nu=1$, shown by the dashed line.  
    }
     \label{fig3}
 \end{figure}

\begin{figure}[H]
     \centering
     \includegraphics[width=\linewidth]{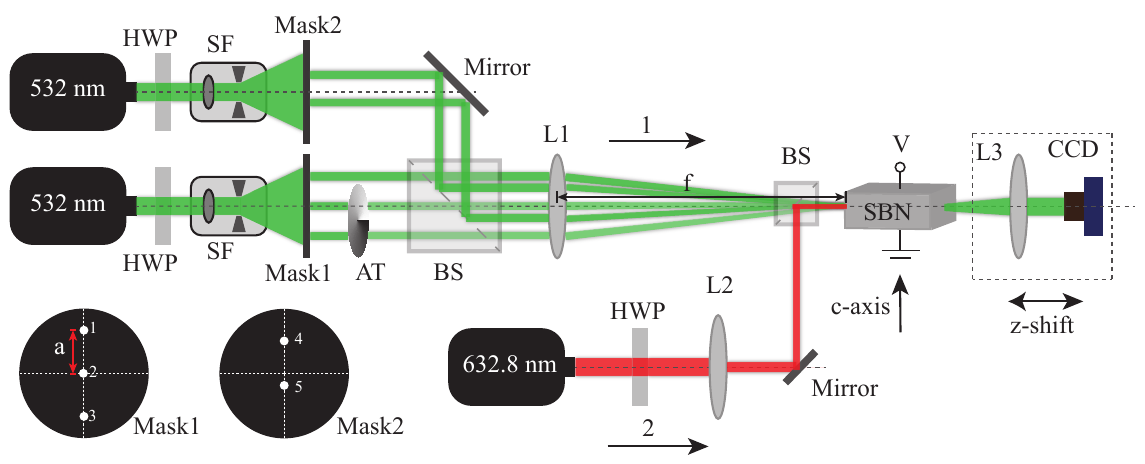}
     \captionsetup{font={stretch=1}}
     \captionsetup{labelfont=bf}
     
     \caption{ {\bf Experimental setup.}
     HWP, half-wave plate; AT, adjustable attenuator; SF, spatial filter; L1, L3, circular lens; L2, cylindrical lens; BS, beam splitter; SBN, strontium barium niobate crystal; CCD, charged-coupled device camera. Inset: Two amplitude masks, one is equipped with three pinholes, and other with two pinholes.    
    }
     \label{fig.4}
\end{figure}

\newpage




\begin{thebibliography}{99}

\bibitem{Thouless1983} 
Thouless, D. J. Quantization of particle transport. \textit{ Phys. Rev. B} \textbf{27}, 6083–6087 (1983). 


\bibitem{Citro23} 
Citro, R. \& Aidelsburger, M. Thouless pumping and topology. \textit{Nat. Rev. Phys.} \textbf{5}, 87 (2023).


\bibitem{Zilberberg2018} 
Zilberberg, O. et al. Photonic topological boundary pumping as a probe of 4D quantum Hall physics. \textit{Nature} \textbf {553}, 59–62 (2018).

\bibitem{Wang2022} 
Wang, P., Fu, Q., Peng, R., Kartashov, Y. V., Torner, L., Konotop, V. V., \&  Ye, F. Two-dimensional Thouless pumping of light in photonic moir\'e lattices. \textit{ Nat. Commun.} \textbf {13}, 6738 (2022).

\bibitem{Benalcazar2022} 
Benalcazar, W. A., Noh, J., Wang, M., Huang, S., Chen, K. P. \& Rechtsman, M. C. Higher-order topological pumping and its observation in photonic lattices. \textit{Phys. Rev. B} \textbf{105}, 195129 (2022).

\bibitem{Cheng2022} 
Cheng, Q., Wang, H., Ke, Y., Chen, T., Yu, Y., Kivshar, Y. S., Lee, C. \& Pan, Y. Asymmetric topological pumping in nonparaxial photonics. \textit{Nat. Commun.} \textbf{13}, 249 (2022).

\bibitem{Ke2016} 
Ke, Y., Qin, X., Mei, F., Zhong, H., Kivshar, Y.S. \& Lee, C. Topological phase transitions and Thouless pumping of light in photonic waveguide arrays. \textit{Laser \& Photon. Rev.} \textbf{10}, 995-1001 (2016).  


\bibitem{Lohse2016} 
Lohse, M., Schweizer, C., Zilberberg, O., Aidelsburger, M. \& Bloch, I., A Thouless quantum pump with ultracold bosonic atoms in an optical superlattice. \textit{Nat. Phys.} \textbf{12}, 350–354 (2016).

\bibitem{Nakajima2016} 
Nakajima, S. et al. Topological Thouless pumping of ultracold fermions. \textit{Nat. Phys.} \textbf{12}, 296–300 (2016).

\bibitem{Taddia2017} 
Taddia, L., Cornfeld, E., Rossini, D., Mazza, L., Sela, E. \& Fazio, R. Topological fractional pumping with Alkaline-earth-like atoms in synthetic lattices. \textit{Phys. Rev. Lett.} \textbf{118}, 230402 (2017).

\bibitem{Lohse2018} 
Lohse, M., Schweizer, C., Price, H. M., Zilberberg, O. \& Bloch, I. Exploring 4D quantum Hall physics with a 2D topological charge pump. \textit{Nature} \textbf{553}, 55–58 (2018).


\bibitem{Ma2018} 
Ma, W., Zhou, L., Zhang, Q., Li, M., Cheng, C., Geng, J., Rong, X., Shi, F., Gong, J. \& Du, J. Experimental observation of a generalized Thouless pump with a single spin. \textit{Phys. Rev. Lett.} \textbf{120}, 120501 (2018).


\bibitem{Rosa2019} 
Rosa, M. I. N., Pal, R. K., Arruda, J. R. F. \& Ruzzene, M. Edge states and topological pumping in spatially modulated elastic lattices. \textit{Phys. Rev. Lett.} \textbf{123}, 034301 (2019).


 \bibitem{Cheng2020} 
 Cheng, W., Prodan, E. \& Prodan, C. Experimental demonstration of dynamic topological pumping across incommensurate bilayered acoustic metamaterials. \textit{Phys. Rev. Lett.} \textbf{125}, 224301 (2020).

\bibitem{Chen2021} 
Chen, H., Zhang, H., Wu, Q., Huang, Y., Nguyen, H., Prodan, E., Zhou, X. \& Huang, G. Creating synthetic spaces for higher-order topological sound transport. \textit{Nat. Commun.} \textbf{12}, 5028 (2021).


\bibitem{Kraus2012} 
Kraus, Y. E., Lahini, Y., Ringel, Z., Verbin, M. \& Zilberberg, O. Topological states and adiabatic pumping in quasicrystals. \textit{Phys. Rev. Lett.} \textbf{109}, 106402 (2012).

\bibitem{Verbin2015} 
Verbin, M., Zilberberg, O., Lahini, Y., Kraus, Y. E. \& Silberberg, Y. Topological pumping over a photonic Fibonacci quasicrystal. \textit{Phys. Rev. B} \textbf{91}, 064201 (2015).

\bibitem{Kraus2013} 
Kraus, Y. E., Ringel, Z. \& Zilberberg, O. Four-dimensional quantum hall effect in a two-dimensional quasicrystal. \textit{Phys. Rev. Lett.} \textbf{111}, 226401 (2013).

\bibitem{Yang2024} 
Yang, K., Fu, Q., Prates, Huang, C., Wang, P., Kartashov, Y. V., Konotop, V. V.\& Ye, F. Observation of Thouless pumping of light in quasi-periodic photonic crystals. \textit{Proc. Natl Acad. Sci. USA} \textbf{121} (2024).

\bibitem{Cerjan2020}
Cerjan, A., Wang, M., Huang, S., Chen, K. P. \& Rechtsman, M. C. Thouless pumping in disordered photonic systems. \textit{Light. Sci. Appl.} \textbf{9}, 178 (2020).



\bibitem{Nakajima2021} 
Nakajima S. et al. Competition and interplay between topology and quasi-periodic disorder in Thouless pumping of ultracold atoms. \textit{Nat. Phys.} \textbf{17},844–849 (2021).


\bibitem{Thouless1982} 
Thouless, D. J., Kohmoto, M., Nightingale, M. P., \& den Nijs, M. Quantized Hall conductance in a two-dimensional periodic potential. \textit{Phys. Rev. Lett.} \textbf{49}, 405 (1982).


\bibitem{Avron1987} 
Avron, J. E., Seiler, R. \& Yaffe, L. G. Adiabatic theorems and applications to the quantum Hall effect. \textit{Com. Math. Phys.} \textbf{110}, 33-49 (1987).


\bibitem{Fu2022a} 
Fu, Q., Wang, P., Kartashov, Y. V., Konotop, V. V. \& Ye, F. Two-dimensional nonlinear Thouless pumping of matter waves. \textit{Phys. Rev. Lett.} \textbf{129}, 183901 (2022).

\bibitem{Schiavoni2003}
Schiavoni, M., Sanchez-Palencia, L., Renzoni, F. \& Grynberg, G. Phase control of directed diffusion in a symmetric optical lattice. \textit{Phys. Rev. Lett.} \textbf{90}, 094101 (2003).

\bibitem{Gommers2005}
Gommers, R., Bergamini, S. \& Renzoni, F. Dissipation induced symmetry breaking in a driven optical lattice. \textit{Phys. Rev. Lett.} \textbf{95}, 073003 (2005).

\bibitem{Grossert2016}
Grossert, C., Leder, M., Denisov, S., Hänggi, P. \& Weitz, M. Experimental control of transport resonances in a coherent quantum rocking ratchet. \textit{Nat. Commun.} \textbf{7}, 10440 (2016).

\bibitem{Gorg2019}
Görg, F., Sandholzer, K., Minguzzi, J., Desbuquois, R., Messer, M. \& Esslinger, T. Realization of density-dependent Peierls phases to engineer quantized gauge fields coupled to ultracold matter. \textit{Nat. Phys.} \textbf{15}, 1161 (2019).

\bibitem{Minguzzi2022}
Minguzzi, J., Zhu, Z., Sandholzer, K., Walter, A. S., Viebahn, K. \& Esslinger, T. Topological Pumping in a Floquet-Bloch Band. \textit{Phys. Rev. Lett.} \textbf{129}, 053201 (2022).


\bibitem{Hatami2016} 
Hatami, H., Danieli, C., Bodyfelt, J. D., \& Flach, S. Quasiperiodic driving of Anderson localized waves in one dimension. \textit{ Phys. Rev. E} \textbf{93}, 062205 (2016).

\bibitem{Zhao2022} 
Zhao, H., Mintert, F., Knolle, J. \& Moessner, R. Localization persisting under aperiodic driving. \textit{Phys. Rev. B} \textbf{105}, L220202 (2022).

\bibitem{Else2020} 
Else, D. V., Ho, W. W. \& Dumitrescu, P. T. Long-Lived interacting phases of matter protected by multiple time-translation symmetries in quasiperiodically driven systems. \textit{Phys. Rev. X} \textbf{10}, 021032 (2020).

\bibitem{Nathan2021} 
Nathan, F., Ge, R., Gazit, S., Rudner, M. \& Kolodrubetz, M. Quasiperiodic Floquet-Thouless energy pump. \textit{Phys. Rev. Lett.} \textbf{127}, 166804 (2021).

 \bibitem{Long2021} 
 Long, D. M., Crowley, P. J. D. \&  Chandran A. Nonadiabatic topological energy pumps with quasiperiodic driving. \textit{Phys. Rev. Lett.} \textbf{126}, 106805 (2021).

\bibitem{Qi2021} 
Qi, Z., Refael, G. \& Peng, Y. Universal nonadiabatic energy pumping in a quasiperiodically driven extended system. \textit{Phys. Rev. B} \textbf{104}, 224301 (2021). 

\bibitem{Kolodrubetz2018} 
Kolodrubetz, M. H., Nathan, F., Gazit, S., Morimoto, T., \& Moore, J. E. Topological Floquet-Thouless energy pump. \textit{Phys. Rev. Lett.} \textbf{120}, 150601 (2018). 

\bibitem{Nathen2021} 
Nathan, F., Ge, R., Gazit, S., Rudner, M., \& Kolodrubetz, M. Quasiperiodic Floquet-Thouless energy pump. \textit{Phys. Rev. Lett.} \textbf{127}, 166804 (2021).

\bibitem{Hu2024} 
Hu, P., Wu, H. W., Xie, P., Zhuo, Y., Sun, W., Sheng, Z., \& Pan, Y. Hearing the dynamical Floquet-Thouless pump of a sound pulse. \textit{Phys. Rev. B} \textbf{110}, 085137 (2024).













\bibitem{Khinchin} 
Khinchin, A. Y. \& Teichmann, T. Continued Fractions. \textit{Physics Today} \textbf{17}, 70–71 (1964).


\bibitem{Ostlund1983} 
Ostlund, S., Pandit, R., Rand, D., Schellnhuber H. J. \& Siggia, E. D. One-dimensional Schr\"{o}dinger equation with an almost periodic potential. \textit{Phys. Rev. Lett.} \textbf{50}, 1873 (1983).

\bibitem{Diener01} 
Diener, R. B., Georgakis, G. A., Zhong, J., Raizen, M. \& Niu, Q. Transitions between extended and localized states in a one-dimensional incommensurate optical lattice. \textit{Phys. Rev. A} \textbf{ 64}, 033416 (2001).

\bibitem{Modugno09} 
Modugno, M. Exponential localization in one-dimensional quasi-periodic optical lattices. \textit{New J. Phys.} \textbf{11} 033023 (2009).

\bibitem{Marra2020} 
Marra, P. \& Nitta, M. Topologically quantized current in quasiperiodic Thouless pumps. \textit{Phys. Rev. Res.} \textbf{2}, 042035(R) (2020).

\bibitem{Zezyulin2022} 
Zezyulin, D. A. \& Konotop, V. V. Localization of ultracold atoms in Zeeman lattices with incommensurate spin-orbit coupling. \textit{Phys. Rev. A} \textbf{105} 063323 (2022).


\bibitem{efremidis2002}
Efremidis, N. K., Sears, S., Christodoulides, D. N., Fleischer, J. W. \& Segev, M. Discrete solitons in photorefractive optically induced photonic lattices. \textit{Phys. Rev. E} \textbf{66}(4), 046602 (2002).

\bibitem{fleischer2003}
Fleischer, J. W., Segev, M., Efremidis, N. K. \& Christodoulides, D. N. Observation of two-dimensional discrete solitons in optically induced nonlinear photonic lattices. \textit{Nature} \textbf{422}, 147-150 (2003).

\bibitem{Fedorova2020} 
Fedorova, Z., Qiu, H., Linden, S. \& Kroha, J. Observation of topological transport quantization by dissipation in fast Thouless pumps. \textit{Nat. Commun.} \textbf{11}, 3758 (2020).


\bibitem{Fu2022} 
Fu, Q., Wang, P., Kartashov, Y. V., Konotop, V. V. \& Ye, F. Nonlinear Thouless pumping: solitons and transport breakdown. \textit{Phys. Rev. Lett.} \textbf{ 128}, 154101 (2022).

\bibitem{Wang2020} 
Wang, P., Zheng, Y., Chen, X., Huang, C., Kartashov, Y. V., Torner, L., Konotop, V. V \& Ye, F. Localization and delocalization of light in photonic moir{\'e} lattices. \textit{Nature} \textbf{577}, 42-46 (2020).

\bibitem{Lederer2008} 
Lederer, F., Stegeman, G. I., Christodoulides, D. N., Assanto, G., Segev, M. \& Silberberg, Y. Discrete solitons in optics. \textit{Phys. Rep.} \textbf{463}, 1-126 (2008).


\bibitem{Alfimov}
Alfimov, G. L., Kevrekidis, P. G., Konotop, V. V. \& Salerno M., Wannier functions analysis of the nonlinear Schr\"odinger equation with a periodic potential. \textit{Phys. Rev. E} \textbf{66}, 046608 (2002).

\bibitem{Kohn}
Kohn W., Analytic properties of Bloch waves and Wannier functions. \textit{Phys. Rev.} {\bf 115}, 809-821 (1959).


 
\end{thebibliography}
\end{document}